\documentclass[11pt,tightenlines,superscriptaddress,floatfixolumn,english,aps,notitlepage,nofootinbib]{revtex4-1}
\usepackage[capitalize]{cleveref}
\usepackage{graphicx}
\usepackage{subfigure}
\usepackage{dcolumn}
\usepackage{bm}
\usepackage{color}

\begin{document}
\title{Novel mechanism for electric quadrupole moment generation in relativistic heavy-ion collisions}
\author{Xin-Li Zhao}
\affiliation{Key Laboratory of Nuclear Physics and Ion-beam Application (MOE), Institute of Modern Physics, Fudan University, Shanghai 200433, China}
\affiliation{Shanghai Institute of Applied Physics, Chinese Academy of Sciences, Shanghai 201800, China}
\affiliation{University of Chinese Academy of Sciences, Beijing 100049, China}
\author{Guo-Liang Ma}
\email[]{glma@fudan.edu.cn}
\affiliation{Key Laboratory of Nuclear Physics and Ion-beam Application (MOE), Institute of Modern Physics, Fudan University, Shanghai 200433, China}
\affiliation{Shanghai Institute of Applied Physics, Chinese Academy of Sciences, Shanghai 201800, China}
\author{Yu-Gang Ma}
\email[]{mayugang@fudan.edu.cn}
\affiliation{Key Laboratory of Nuclear Physics and Ion-beam Application (MOE), Institute of Modern Physics, Fudan University, Shanghai 200433, China}
\affiliation{Shanghai Institute of Applied Physics, Chinese Academy of Sciences, Shanghai 201800, China}


\begin{abstract}
We present the spatial distributions of electromagnetic fields ($\bf E$ and $\bf B$) and electromagnetic anomaly $ \bf E \cdot B$ in Au+Au collisions at the RHIC energy $\sqrt{s}$=200 GeV based on a multi-phase transport model.  A dipolar distribution of $\bf E \cdot B$ is observed in non-central collisions. We find that the coupling of the $\bf E \cdot B$ dipole and magnetic field $\bf B$ can induce an electric quadrupole moment which can further lead to the difference in elliptic flows between positive charged particles and negative charged particles through final interactions. The centrality dependence of the density of $\bf E \cdot B$ is similar to the trend of the slope parameter $r$ measured from the difference in elliptic flows between positive pions and negative pions by the STAR collaboration. Therefore, the novel mechanism for electric quadrupole moment generation can offer a new interpretation of the observed charge-dependent elliptic flow of pions, but without the formation of chiral magnetic wave.

\end{abstract}


\maketitle

\section{Introduction}
\label{introduction}

In non-central heavy-ion collisions, the fast-moving charged ions can produce a strong magnetic field with the magnitude about $eB \sim m_\pi^2 \sim 10^{18}$ to  $10^{19}$ Gauss from the energy available at the BNL Relativistic Heavy Ion Collider (RHIC) to the energy available at the CERN Large Hadron Collider (LHC)~\cite{Bzdak:2011yy,Deng:2012pc,Zhao:2017rpf}. Meanwhile, it is believed that relativistic heavy-ion collisions at RHIC and LHC produce a quark-gluon plasma (QGP), in which some chiral anomalous transport effects could happen~\cite{Kharzeev:2015znc,Fukushima:2008xe,Kharzeev:2004ey,Kharzeev:2007tn,Hattori:2016emy}. For instance, in the presence of an external magnetic field, if a chiral anomaly occurs in QGP which leads to a nonzero axial chemical potential $\mu_A$, it will induce a vector current ${\bf J}_V$ (Chiral Magnetic Effect)~\cite{Kharzeev:2007jp,Fukushima:2008xe,Kharzeev:2013ffa,Kharzeev:2015znc,Huang:2015oca} :
\begin{eqnarray}
{{\bf J}_V}=\frac{Qe}{2\pi ^{2}}\mu _{A}{\bf B}.
\label{JV}
\end{eqnarray}%

The axial charge chemical potential $\mu_A$ is believed to be related to chiral anomaly in a plasma with chiral fermions. It is believed that there are three main sources to generate local axial charges, and the axial anomaly equation with the inclusion of mass term reads~\cite{Iatrakis:2015fma,Guo:2016nnq}
\begin{eqnarray}
\partial_{\mu}J_{5}^{\mu}=2im\bar{\psi}\gamma^{5}\psi-\frac{e^2}{16\pi^2}\epsilon^{\mu\nu\rho\sigma}F_{\mu\nu}F_{\rho\sigma}-\frac{g^2}{16\pi^2}tr\epsilon^{\mu\nu\rho\sigma}G_{\mu\nu}G_{\rho\sigma},
\label{DJA}
\end{eqnarray}%
where the first term on the right hand side is a source coming from non-zero quark mass due to chiral symmetry breaking, the second and third terms corresponds to the QED and QCD anomalies, respectively.  In this work, we focus on the QED anomaly, which is usually represented by the second Lorentz invariant $I_2=\bf E \cdot \bf B$, in relativistic heavy-ion collisions.

Similarly to the CME, with the external magnetic field, a nonzero vector chemical potential $\mu_V$ will induce an axial current ${\bf J}_A$(Chiral Separation Effect)~\cite{Son:2004tq,Metlitski:2005pr}:
\begin{eqnarray}
{{\bf J}_A}=\frac{Qe}{2\pi ^{2}}\mu _{V}{\bf B}.
\label{JA}
\end{eqnarray}%

Therefore, the phenomena of electric charge separation and chiral charge separation along the magnetic field direction can happen for the CME and CSE, respectively. With the presence of magnetic field, the coupling of the vector and axial currents induced by the two potentials can motivate a collective gapless excitation in QGP, i.e. the Chiral Magnetic Wave (CMW)~\cite{Kharzeev:2010gd,Burnier:2012ae,Yee:2013cya}. An analogy to electromagnetic wave, one can understand the CMW by a simple intuitive way: first a nonzero vector chemical potential $\mu_V$ induces a CSE axial current ${\bf J}_A$ which can lead to a dipolar separation of axial charges which are along the magnetic $\bf B$ field direction; next the nonzero axial chemical potential $\mu_A$ in turn induces a CME current ${\bf J}_V$ in the opposite direction; then the nonzero vector and axial charge density mutually induce each other resulting in a charge quadrupole that there are more positive charges at the poles of almond-shape fireball (since $\mu_V > 0$ and $\bf B$ is primarily out-of-plane) than at the equator (in the reaction plane) of the fireball of the QGP~\cite{Burnier:2011bf,Gorbar:2011ya}.  The charge quadrupole distribution can further lead to the difference in elliptic flows between positive charged particles and negative charged particles through final interactions~\cite{Burnier:2011bf,Ma:2014iva}.  The elliptic flow $v_{2}$ therefore becomes charge-dependent:

\begin{eqnarray}
v_{2}^{\pm}=v_{2,\pm}^{base}\mp r A_{ch}/2
\label{deltav2}
\end{eqnarray}%

where $A_{ch}=\frac{N_{+}-N_{-}}{N_{+}+N_{-}}$ is the charge asymmetry, and the slope parameter $r$ can be used to parametrize reflecting the strength of the asymmetry in the azimuthal distributions of positive and negative hadrons, which should be proportional to the electric quadrupole moment. The recent STAR measurements seem to support the existence of the CMW~\cite{Adamczyk:2015eqo}, though it is still on debate~\cite{Hatta:2015hca,Bzdak:2013yla}. 

From Eq.~(\ref{DJA}), we know the QED anomaly is one important source of the chiral anomaly, which indeed can play a key role in some chiral anomalous effects. For instance, Kharzeev argued that the field configuration of parallel electric $\bf E$ fields and magnetic $\bf B$ fields will skew the balance between the Fermi surfaces of left- and right-handed fermions in the Dirac sea~\cite{Kharzeev:2009fn,Kharzeev:2013ffa}. According to Eq.~(\ref{JV}), we know the CME can happen and induce a vector current ${\bf J}_V$, since $\mu _{A} \propto\bf E \cdot \bf B$. In addition, Cao and Huang introduced the nonzero second Lorentz invariant $I_2=\bf E \cdot \bf B$ which provides a parity-odd environment in which the otherwise-forbidden neutral pion condensation can occur via the electromagnetic triangle anomaly~\cite{Cao:2015cka}. With parallel electric and magnetic fields, the CME has been experimentally observed in some Dirac semimetals~\cite{Li:2014bha,Huang:2015eia,Shekhar:2015rqa}. In this work, we will turn out that the spatial distribution of $\bf E \cdot \bf B$ in the transverse plane is a configuration with a dipolar shape. What's more, the coupling of the $\bf E \cdot \bf B$ dipole and magnetic field $\bf B$ can also produce an electric quadrupole moment. That means that as long as there is an $\bf E \cdot \bf B$ dipole in an external magnetic field, an electric quadrupole moment can also be generated without the formation of the CMW. We will describe this phenomenon in detail in the following sections.

The paper is organized as follows. In Sec.~\ref{GS}, we provide a brief introduction to a multi-phase transport (AMPT) model, our methods to calculate electromagnetic fields and the density of $\bf E \cdot \bf B$. The numerical results for the properties of electromagnetic fields and $\bf E \cdot \bf B$ are presented and discussed in details in Sec.~\ref{results}. Sec.~\ref{summary} contains our conculsions.

\section{GENERAL SETUP}
\label{GS}
\subsection{AMPT model}

In this work, the AMPT model with a string melting scenario is utilized~\cite{Lin:2004en}. The AMPT model is a hybrid Monte Carlo model, invented to simulate relativistic heavy-ion collisions which can well describe many experimental observables for Au+Au collisions at RHIC energies and Pb+Pb collisions at LHC energies~\cite{Zhang:1999bd,Bass:1999zq,Lin:2000cx}. At present, there exist two versions of AMPT model, the default version and the version with a string melting mechanism. Both versions consist of four important evolution stages of heavy-ion collisions: initial state, parton cascade, hadronization, and hadron rescatterings. The initial state of collisions is generated by using the HIJING model~\cite{Wang:1991hta,Gyulassy:1994ew}. The difference between the two versions is that in the string-melting version, strings and minijets are melted into partons, so that there are more partons participating in parton cascade than the default version. Therefore, the string-melting version can better describe the cases when the QGP is produced, such as heavy-ion collisions at RHIC and LHC energies. In our convention, we choose $x$ axis along the direction of impact parameter $b$ from the target center to the projectile center, $z$ axis along the beam direction, and $y$ axis perpendicular to the $x$ and $z$ directions, by using the string-melting version of AMPT model.

\subsection{Calculations of electromagnetic field}

Following Refs.~\cite{Deng:2012pc,Zhao:2017rpf}, we use the same way to calculate the initial electromagnetic fields as follows,
\begin{eqnarray}
e{\bf E}(t,{\bf r})&=&\frac{e^2}{4\pi}{\sum\limits_{n}}Z_{n} \frac{{\bf R}_n- R_n{\bf v}_n}{(R_n-{\bf R}_n \cdot {\bf v}_n)^3}(1-v_n^2),
\label{elec}\\
e{\bf B}(t,{\bf r})&=&\frac{e^2}{4\pi}{\sum\limits_{n}}Z_{n} \frac{{\bf v}_n \times {\bf R}_n}{(R_n-{\bf R}_n \cdot {\bf v}_n)^3}(1-v_n^2),
\label{magn}
\end{eqnarray}%
where we use natural unit $\hbar = c = 1$, $Z_n$ is the charge number of the $n$th particle, for proton it is one, ${\bf R}_n = {\bf r} - {\bf r}_n$ is the relative position of the field point {\bf r} to the source point ${\bf r}_n$, and ${\bf r}_n$ is the location of the $n$th particle with velocity ${\bf v}_n$ at the retarded time $t_{n} = t - |{\bf r} - {\bf r}_n|$. The summations run over all charged protons in the system. We need to emphasis that most of our results about their electromagnetic fields are calculated at the field point ${\bf r}$= (0, 0, 0) at $t$ = 0. Then we can calculate $\bf E \cdot \bf B$ by using the following formula,
\begin{eqnarray}
{\bf E} \cdot {\bf B} = E_x B_x + E_y B_y + E_z B_z.
\label{EdotB}
\end{eqnarray}%

Because the longitudinal electromagnetic fields are much smaller than the transverse fields~\cite{Bzdak:2011yy,Deng:2012pc,Zhao:2017rpf},  we neglect the term of $E_z B_z$, i.e., we only investigate $\bf E \cdot \bf B$ for the transverse plane. It is emphasized that we calculate $\bf E \cdot \bf B$ on the event-by-event basis, which means that  $\langle \bf E \cdot B\rangle$ represents the average of events, i.e. the bracket $\langle ... \rangle$ means taking average over all events in the following sections.

\subsection{Calculations of the density of $\langle \bf E \cdot B\rangle$}

\begin{figure*}[htb]
  \setlength{\abovecaptionskip}{0pt}
  \setlength{\belowcaptionskip}{8pt}\centerline{\includegraphics[scale=0.5]{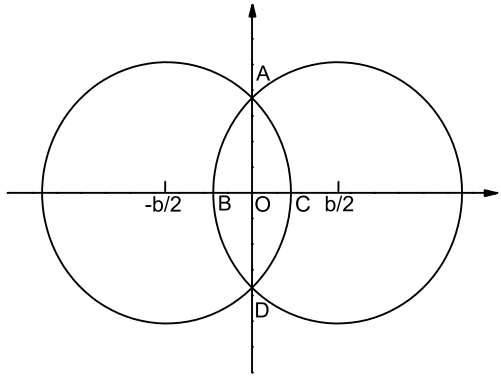}}
\caption{The geometrical illustration of non-central collisions with an impact parameter $b$.}\label{fig:x0y}
\end{figure*}

In this subsection, we describe how to calculate the density of $\langle \bf E \cdot B\rangle$. First, we need to calculate the collision area in the collision region of two nuclei, as shown in Fig.
~\ref{fig:x0y}. However, it is difficult to determine the extent of the collision area. Here we use four following methods to calculate the collision area. The main difference among these four methods is coming from the nuclear radius r$_{A}$, because once we know r$_{A}$, the collision region of two nuclei can be calculated by the ideal geometry intersection region as shown in Fig.~\ref{fig:x0y}. The r$_{A}$ for the four methods is described in the following: the first method (denoted as M1 thereafter) sets r$_{A}$ = 6.5 fm; the second method (denoted as M2 thereafter) sets r$_{A}$ = 7 fm; the third method (denoted as M3 thereafter) is that, firstly the length of OC segment is determined by the distribution of initial partons based on the AMPT model, then the radius of nucleus r$_{A}$ is determined by OC segment according to the ideal geometric relation of the collision region; the fourth method (denoted as M4 thereafter) is similar to M3, except that firstly the length of OA segment is determined by the distribution of initial partons, then the radius of nucleus r$_{A}$ is determined by OA segment according to the ideal geometric relation of the collision region. Then, after we caculate the distribution of $\langle \bf E \cdot B\rangle$ in the collision area, it is not difficult get the density of $\langle \bf E \cdot B\rangle$.

\section{Results and discussions}
\label{results}

\subsection{Spatial distributions of electromagnetic fields}
\label{resultsA}

\begin{figure*}[htb]
  \setlength{\abovecaptionskip}{0pt}
  \setlength{\belowcaptionskip}{8pt}\centerline{\includegraphics[scale=0.6]{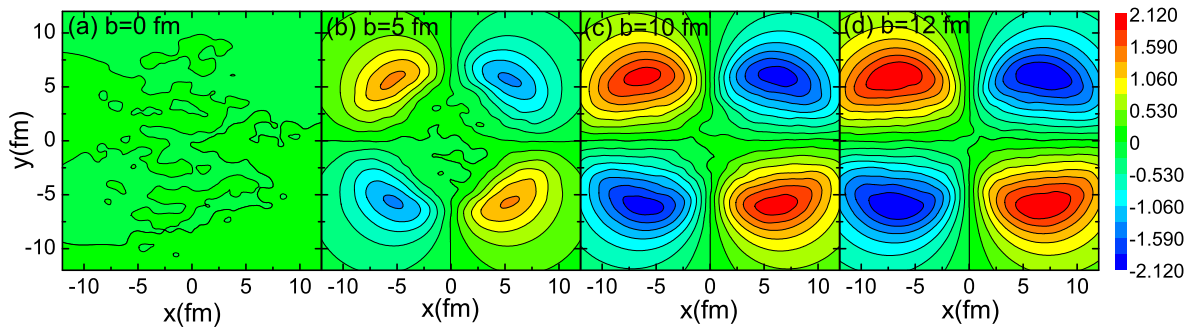}}
\caption{(Color online) The spatial distributions of $e\langle B_{x}\rangle$ in the transverse plane at $t = 0$ for (a) $b = 0$ fm (b) $b = 5$ fm (c) $b = 10$ fm (d) $b = 12$ fm in Au+Au collisions at $\sqrt s = 200$ GeV, where the unit is $m_\pi^2$.}\label{fig:Bx}
\end{figure*}
\begin{figure*}[htb]
  \setlength{\abovecaptionskip}{0pt}
  \setlength{\belowcaptionskip}{8pt}\centerline{\includegraphics[scale=0.6]{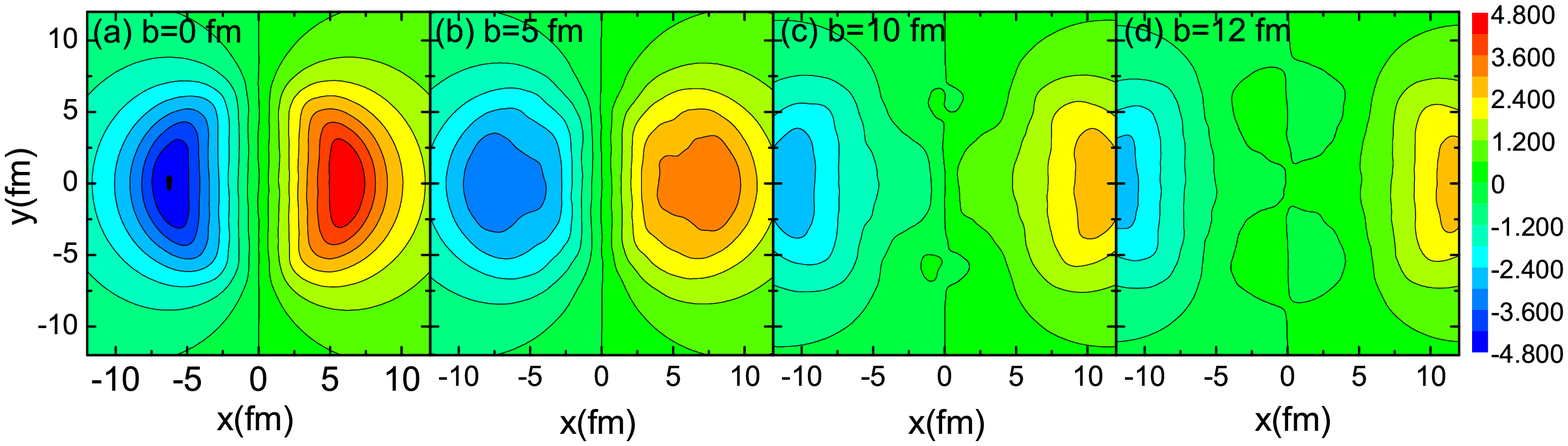}}
\caption{(Color online) The spatial distributions of $e\langle E_{x}\rangle$ in the transverse plane at $t = 0$ for (a) $b = 0$ fm (b) $b = 5$ fm (c) $b = 10$ fm (d) $b = 12$ fm in Au+Au collisions at $\sqrt s = 200$ GeV, where the unit is $m_\pi^2$.}\label{fig:Ex}
\end{figure*}
\begin{figure*}[htb]
  \setlength{\abovecaptionskip}{0pt}
  \setlength{\belowcaptionskip}{8pt}\centerline{\includegraphics[scale=0.6]{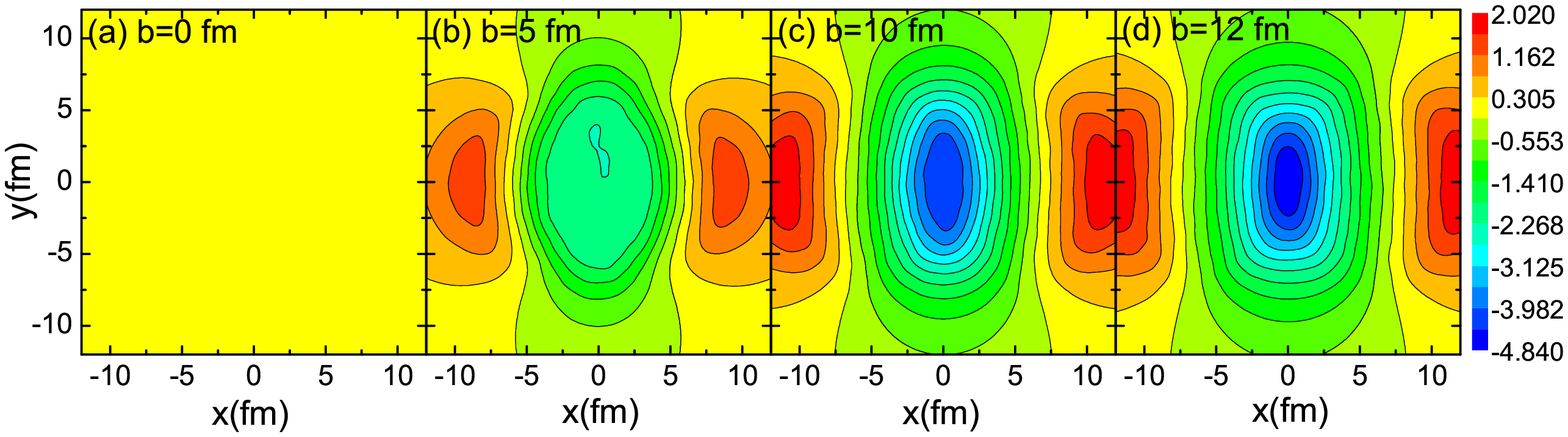}}
\caption{(Color online) The spatial distributions of $e\langle B_{y}\rangle$ in the transverse plane at $t = 0$ for (a) $b = 0$ fm (b) $b = 5$ fm (c) $b = 10$ fm (d) $b = 12$ fm in Au+Au collisions at $\sqrt s = 200$ GeV, where the unit is $m_\pi^2$.}\label{fig:By}
\end{figure*}
\begin{figure*}[htb]
  \setlength{\abovecaptionskip}{0pt}
  \setlength{\belowcaptionskip}{8pt}\centerline{\includegraphics[scale=0.6]{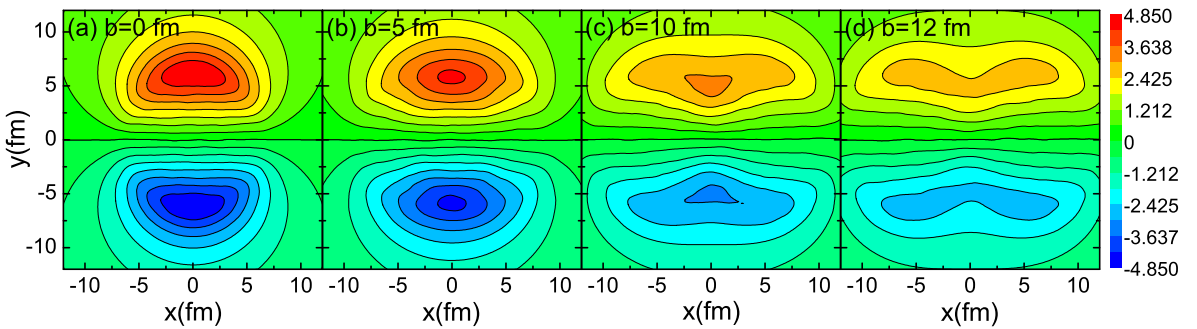}}
\caption{(Color online) The spatial distributions of $e\langle E_{y}\rangle$ in the transverse plane at $t = 0$ for (a) $b = 0$ fm (b) $b = 5$ fm (c) $b = 10$ fm (d) $b = 12$ fm in Au+Au collisions at $\sqrt s = 200$ GeV, where the unit is $m_\pi^2$.}\label{fig:Ey}
\end{figure*}

$\langle \bf E \cdot B\rangle$ is mainly affected by the transverse components of the magnetic field and electric field. In Figs.~\ref{fig:Bx}-\ref{fig:Ey}, we show the contour plots of $\langle B_{x}\rangle$, $\langle E_{x}\rangle$, $\langle B_{y}\rangle$ and $\langle E_{y}\rangle$ at $t$ = 0 in the transverse plane for Au+Au collisions at $\sqrt{s}$ = 200 GeV. We know the fields depend on centrality, so we show the contour plots of $\langle B_{x}\rangle$, $\langle E_{x}\rangle$, $\langle B_{y}\rangle$ and $\langle E_{y}\rangle$ for different centralities, i.e. the plots (a)-(d) in Figs.~\ref{fig:Bx}-\ref{fig:Ey} show $b$ = 0, 5, 10 and 12 fm, respectively. It is easy to see the spatial distributions of electromagnetic fields are very inhomogeneous for different centralities. Moreover, it should be mentioned that our results of $\langle B_{x, y}\rangle$ and $\langle E_{x, y}\rangle$ at $b$= 0 and 10 fm are very similar to those in Ref.~\cite{Deng:2012pc}.

\subsection{Spatial distributions of $\langle \bf E \cdot B\rangle$}
\label{resultsB}

\begin{figure*}[htb]
  \setlength{\abovecaptionskip}{0pt}
  \setlength{\belowcaptionskip}{8pt}\centerline{\includegraphics[scale=0.6]{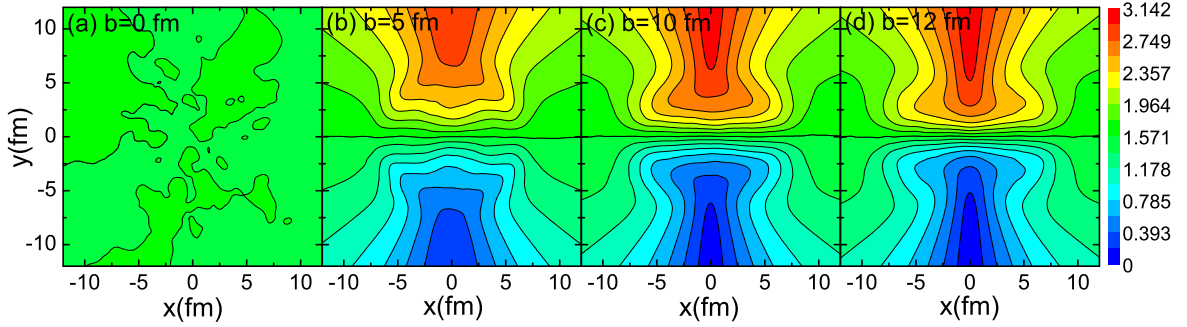}}
\caption{(Color online) The spatial distributions of the angle between electric and magnetic fields in the transverse plane at $t = 0$ for (a) $b = 0$ fm (b) $b = 5$ fm (c) $b = 10$ fm (d) $b = 12$ fm in Au+Au collisions at $\sqrt s = 200$ GeV.}\label{fig:cosEB}
\end{figure*}

Fig.~\ref{fig:cosEB} shows the relative angle between magnetic field and electric field for different centralities. At most central collisions, the angles are almost $\pi/2$, which means $\langle \bf E \cdot B\rangle$ = 0 corresponding to $\mu_A$ = 0 and no chiral anomaly happens. However, for non-central collisions, the direction of the magnetic field is parallel or antiparallel to that of the electric field somewhere (where the angles can be 0 or $\pi$), which means $\langle \bf E \cdot B\rangle \neq$ 0 corresponding to $\mu_A \neq 0$ and chiral anomaly happens.

\begin{figure*}[htb]
  \setlength{\abovecaptionskip}{0pt}
  \setlength{\belowcaptionskip}{8pt}\centerline{\includegraphics[scale=0.6]{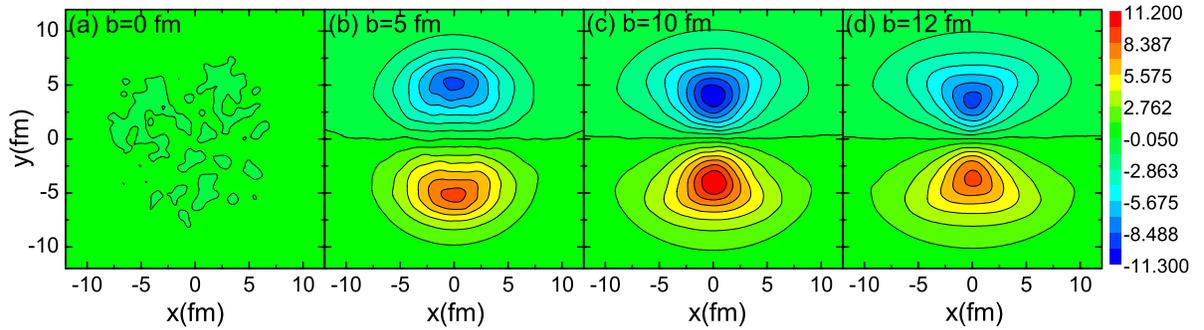}}
\caption{(Color online) The spatial distributions of $e^2\langle \bf E \cdot B\rangle$ in the transverse plane at $t = 0$ for (a) $b = 0$ fm (b) $b = 5$ fm (c) $b = 10$ fm (d) $b = 12$ fm in Au+Au collisions at $\sqrt s = 200$ GeV, where the unit is $m_\pi^4$.}\label{fig:EBxy}
\end{figure*}

\begin{figure*}[htb]
  \setlength{\abovecaptionskip}{0pt}
  \setlength{\belowcaptionskip}{8pt}
  \centerline{
  \includegraphics[scale=0.3]{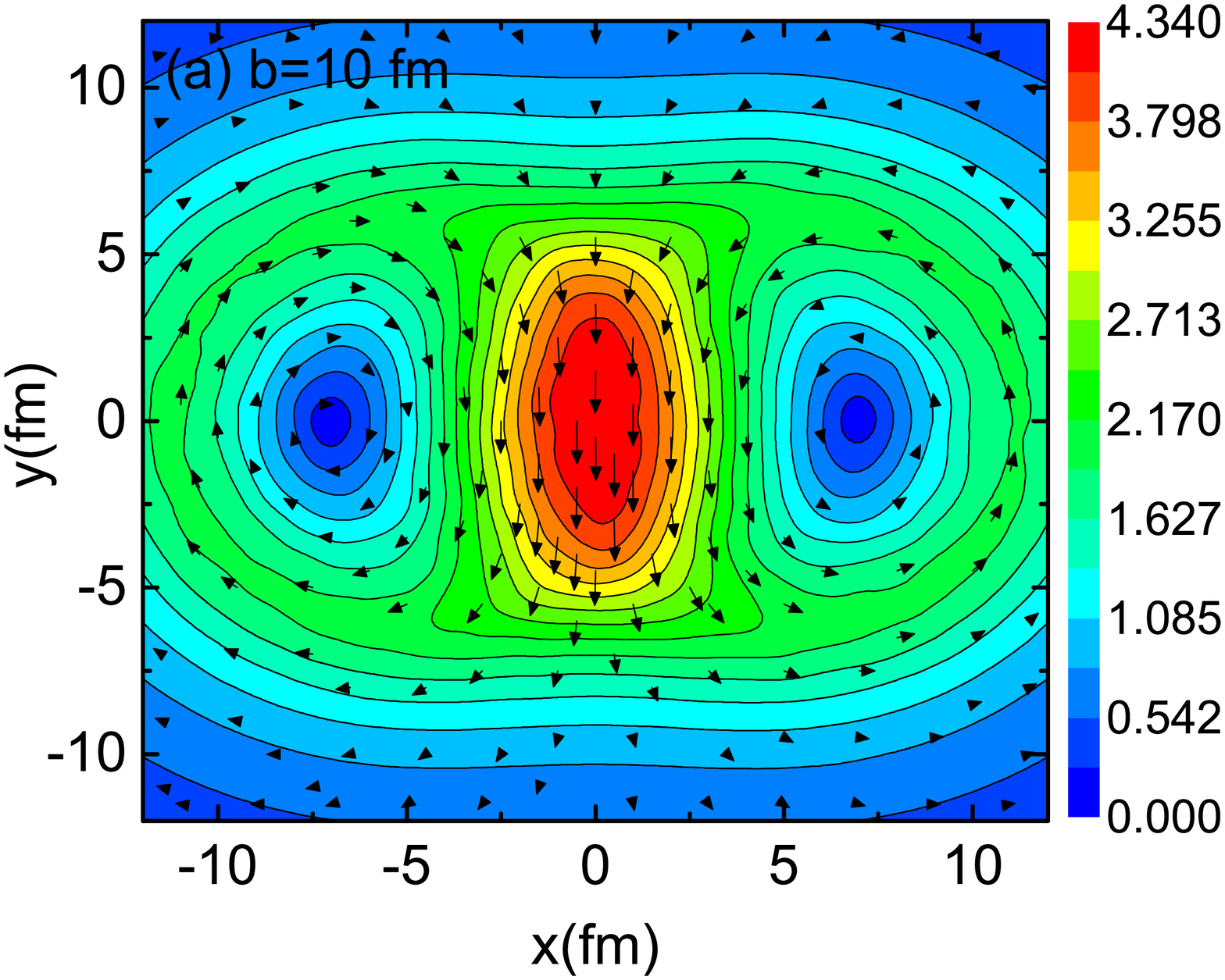}
  \includegraphics[scale=0.3]{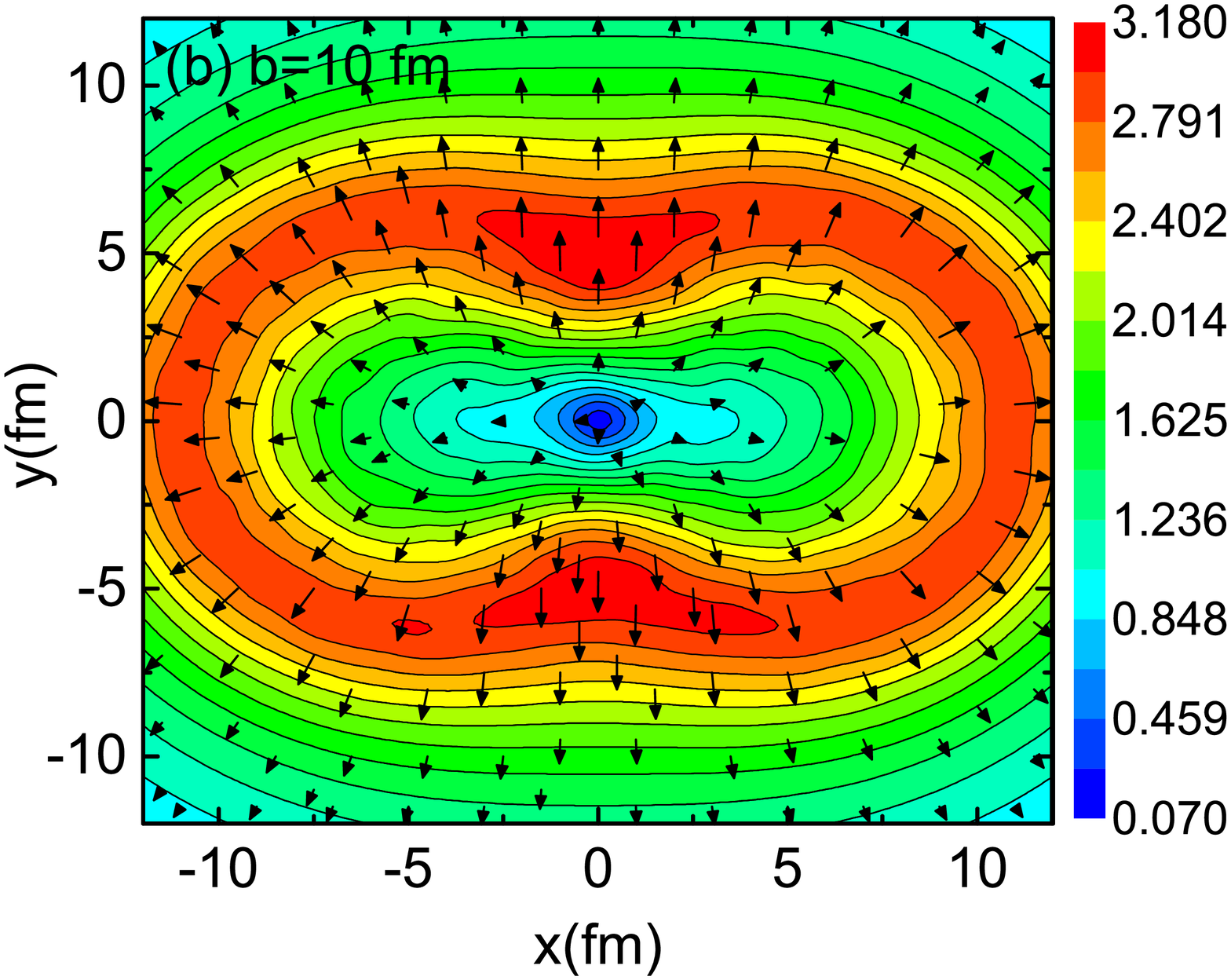}
}
\caption{(Color online) The spatial distributions of total $\bf B$ (a) and $\bf E$ (b) fields with magnetic and electric field lines (where arrows denote the field directions) in the transverse plane at $t = 0$ for $b = 10$ fm in Au+Au collisions at $\sqrt s = 200$ GeV, where the unit is $m_\pi^2$.}\label{fig:EBT}
\end{figure*}

\begin{figure*}[htb]
  \setlength{\abovecaptionskip}{0pt}
  \setlength{\belowcaptionskip}{8pt}\centerline{\includegraphics[scale=0.6]{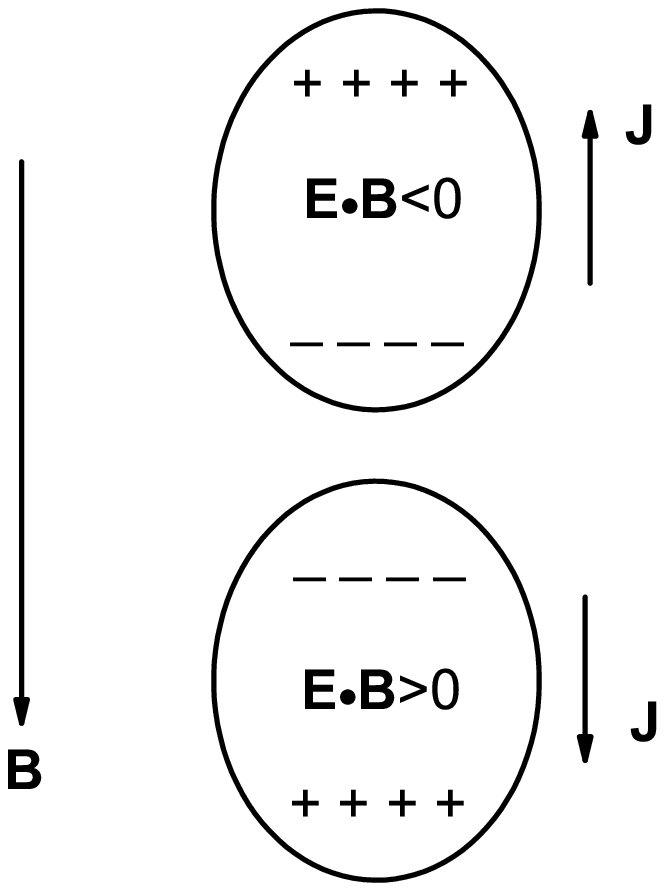}}
\caption{The illustration of the coupling of $\langle \bf E \cdot B\rangle$ and $\bf B$ producing an electric quadrupole moment in Au+Au collisions at $\sqrt s = 200$ GeV for non-central collisions.}\label{fig:cmw}
\end{figure*}

Fig.~\ref{fig:EBxy} shows the spatial distributions of $\langle \bf E \cdot B\rangle$ in the transverse plane at $t$ = 0 for different centralities, where (a)-(d) are $b$ = 0, 5, 10 and 12 fm, respectively.  For most central collisions,  $\langle \bf E \cdot B\rangle$ fluctuates, but it is close to zero everywhere. However, for non-central collisions, the spatial distribution of $\langle \bf E \cdot B\rangle$ is analogous to a dipole, which $\langle \bf E \cdot B\rangle < 0$ in the positive transverse plane ($y > 0$) but $\langle \bf E \cdot B\rangle > 0$ in the negative transverse plane ($y < 0$). Fig.~\ref{fig:EBT} shows an example how the magnetic and electric field lines distributed in the transverse plane for total $\bf B$ and $\bf E$ at $t = 0$ for peripheral Au+Au collisions ($b = 10$ fm), through which ones can easily understand the cause of the dipole. This dipole will lead to a very interesting consequence. According to Eq.~(\ref{JV}), Non-zero $\langle \bf E \cdot B\rangle$ (or $\mu_A$) can lead to a CME current in the presence of magnetic field. As shown in Fig.~\ref{fig:cmw}, the existence of current $\bf J$ will finally induce a electric quadruple moment,  which is similar to the CMW-induced electric quadruple moment. In this sense, it means that the electromagnetic anomaly can also induce an electric quadruple moment in the QGP but without the formation of CMW. It shows a same charged configuration with the 'equator' accumulating additional negative charges and the two poles (out-of-plane) accumulating additional positive charges. It also can give rise to the difference in elliptic flows between positive charged particles and the negative charged particles with the help of final state interactions. We have checked that the two CME current directions have no change if we switch the directions of velocity of colliding nuclei, because the direction of magnetic field and the sign of $\langle \bf E \cdot B\rangle$ will also change.

\begin{figure*}[htb]
  \begin{minipage}[t]{0.333\linewidth}
    \includegraphics[width=0.95\textwidth]{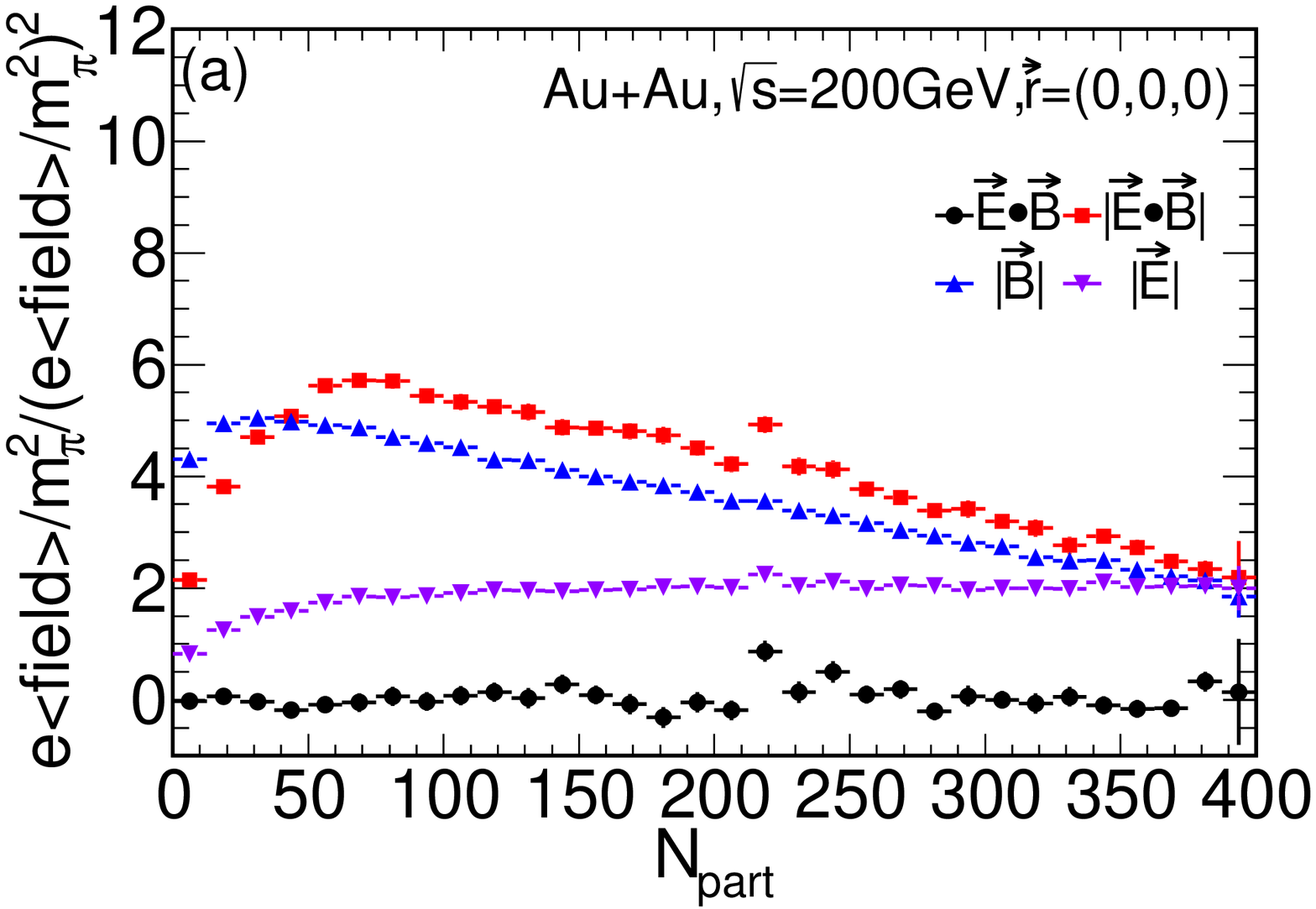}
    \label{fig:side:a}
  \end{minipage}%
  \begin{minipage}[t]{0.333\linewidth}
    \centering
    \includegraphics[width=0.95\textwidth]{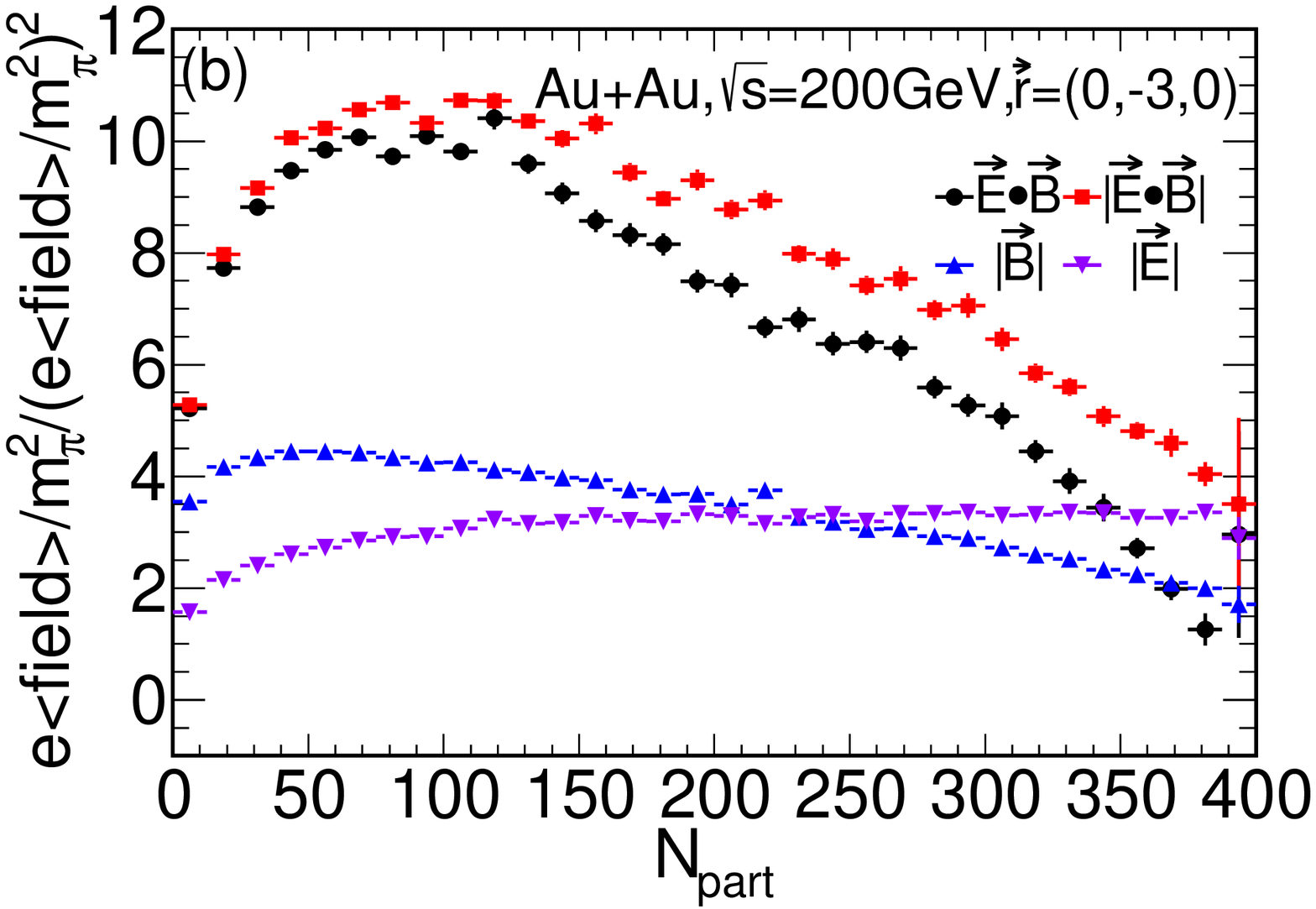}
    \label{fig:side:b}
  \end{minipage}%
  \begin{minipage}[t]{0.333\linewidth}
    \centering
    \includegraphics[width=0.95\textwidth]{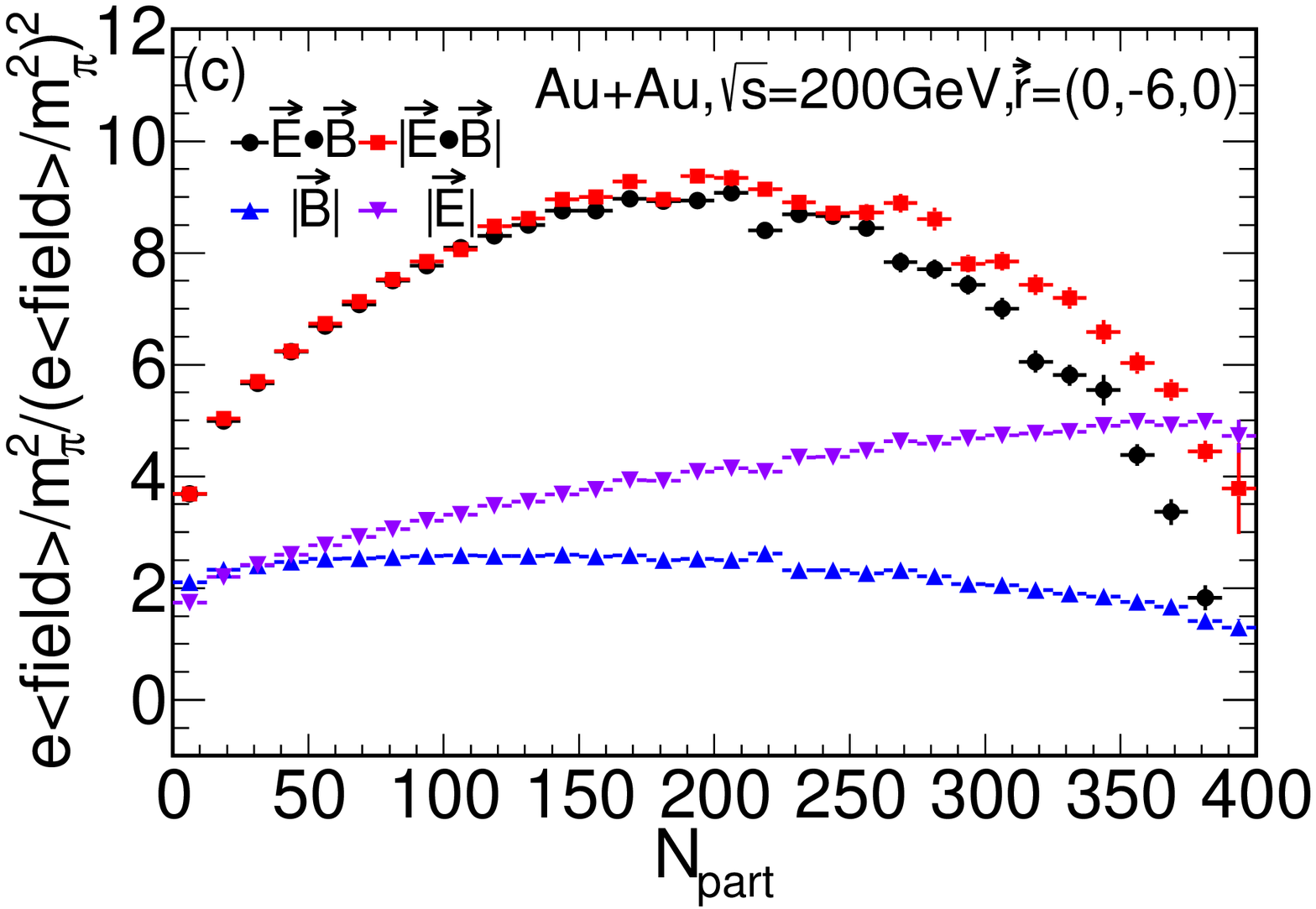}
    \label{fig:side:b}
  \end{minipage}
\caption{(Color online) The electromagnetic fields, $\langle \bf E \cdot B\rangle$ and $\langle |\bf E \cdot B|\rangle$ as functions of $N_{part}$ at $t = 0$ and field points (a) ${\bf r}=(0, 0, 0)$ (b) ${\bf r}=(0, -3 fm, 0)$ (c) ${\bf r}=(0, -6 fm, 0)$ in Au+Au collisions at $\sqrt s = 200$ GeV.}
\label{fig:EBpart}
\end{figure*}

From Fig.~\ref{fig:EBxy}, we see that the spatial distribution of $\langle \bf E \cdot B\rangle$ is inhomogeneous. To study these inhomogeneities, Fig.~\ref{fig:EBpart} shows $N_{part}$ dependences of the electromagnetic fields and their anomaly at three selected different field points [${\bf r}=(0, 0, 0)$, ${\bf r}=(0, -3, 0)$ and ${\bf r}=(0, -6, 0)$]. It's not difficult to find that the absolute value of magnetic field increases and then gradually decreases as $N_{part}$ increases, but the absolute value of electric field gradually increases as $N_{part}$ increases. Moreover, it is easy to see that the $\langle \bf E \cdot B\rangle \neq \langle |\bf E \cdot B|\rangle$ at field point ${\bf r}=(0, 0, 0)$ indicating large fluctuations of $\bf E \cdot B$. However, $\langle \bf E \cdot B\rangle$ and $\langle |\bf E \cdot B|\rangle$ have similar to trend but not exactly the same at field points ${\bf r}=(0, -3, 0)$ and ${\bf r}=(0, -6, 0)$, indicating some weak fluctuations there. 

\begin{figure*}[htb]
  \begin{minipage}[t]{0.45\linewidth}
    \centering
    \includegraphics[width=0.94\textwidth]{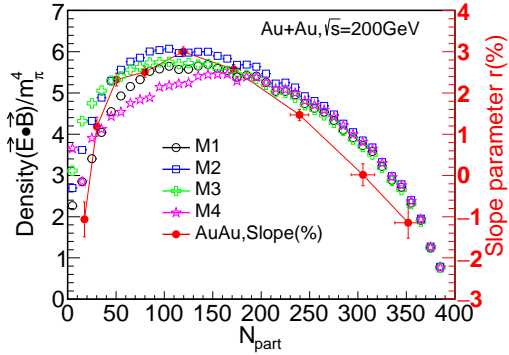}
    \label{fig:side:b}
  \end{minipage}%
\caption{(Color online) The zone-averaged density of $\langle \bf E \cdot B\rangle$ from different calculation methods (open symbols) at $t = 0$ in the transverse plane of $y<0$ fm and the slope parameter (red filled symbol) as functions of $N_{part}$ in Au+Au collisions at $\sqrt s = 200$ GeV.}
\label{fig:denEB}
\end{figure*}

Fig.~\ref{fig:denEB} shows $N_{part}$ dependences of the zone-averaged density of $\langle \bf E \cdot B\rangle$ for four methods M1, M2, M3 and M4, respectively, in comparison with the slope parameter $r$ which is the coefficient reflecting the measured strength of the elliptic flow asymmetry of positive and negative hadrons (or corresponding to the strength of electric quadruple moment). Because of the dipolar symmetry, we choose the lower part of transverse plane ($y<0$ fm) to calculate the averaged value. On the other hand, the density of $\langle \bf E \cdot B\rangle$ can reflect the strength of electric quadruple moment from our point of view. We find that $N_{part}$ dependences of the density of $\langle \bf E \cdot B\rangle$ obtained by the four methods show a similar tendency to the measured trend of the slope parameter $r$. It indicates that a similar effect to the CMW can be induced when such a dipolar configuration of chiral anomaly due to $\langle \bf E \cdot B\rangle$ exists in a large magnetic field. It is interesting to point out that this new mechanism does not need a finite baryon density (i.e. $\mu_{V}=\mu_{B}\neq0$), but which is necessary to drive a chiral separation effect in the CMW mechanism. We expect that our new mechanism could result in a different energy dependence of slope parameter $r$ from the CMW-driven one, which in principle can be examined in the Beam Energy Scan program at RHIC.

\subsection{Spatial distributions of $\langle \bf E \cdot B\rangle$ for a single event}
\label{resultsC}

\begin{figure*}[htb]
  \setlength{\abovecaptionskip}{0pt}
  \setlength{\belowcaptionskip}{8pt}\centerline{\includegraphics[scale=0.63]{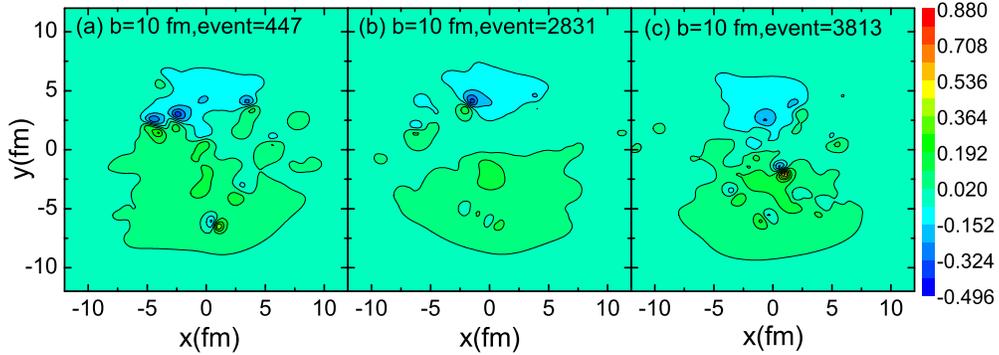}}
\caption{(Color online) The spatial distributions of $e^2 \bf E \cdot B$ in the transverse plane at $t = 0$ for $b = 10$ fm in Au+Au collisions at $\sqrt s = 200$ GeV from three randomly selected events, where the unit is $m_\pi^4\frac{}{}$.}\label{fig:eventEB}
\end{figure*}

However, someone may think our results are event averaged results, because there could be some events in which the configuration of $\langle \bf E \cdot B\rangle$ could be completely opposite. Fig.~\ref{fig:eventEB} shows the spatial distributions of $\bf E \cdot B$ for three randomly-selected single events. We can find that for single event distribution of $\bf E \cdot B$, it is still negative in the positive-y half-plane, but positive in the negative-y half-plane at $b$ = 10 fm for three randomly-selected single events, though there are some fluctuations. This supports that the spatial distribution of $\langle \bf E \cdot B\rangle$, as shown in Fig.~\ref{fig:EBxy},  does work not only for event averaged but also for a single event. In principle, it can provide an important chiral anomalous environment to make some chiral anomalous transport effects occur even without any QCD topology-changing transitions.

\section{Conclusions}
\label{summary}
We investigate the spatial distributions of electromagnetic fields and their anomaly of $\bf E \cdot B$ in Au+Au collisions at the RHIC energy $\sqrt{s}$=200 GeV by using AMPT model. A dipolar distribution of $\bf E \cdot B$ is observed inside the fireball, which could provide an chiral anomalous environment to make some chiral anomalous transport effects occur. We argue that the coupling of dipolar $\bf E \cdot B$ distribution and magnetic field $\bf B$ can induce an electric quadrupole moment which can further give rise to the difference in elliptic flows between positive charged particles and the negatively charged particles. Our calculations demonstrate that the centrality dependence of the density of $\bf E \cdot B$ is consistent with the measured dependence of the slope parameter $r$ by the STAR collaboration. Therefore, the observable splitting of elliptic flows of positive and negative pions may not be a necessary evidence for the chiral magnetic wave. Our novel mechanism offers a new possible interpretation of the charge-dependent elliptic flow of pions in Au+Au collisions at RHIC energies observed in the STAR experiment.

\begin{acknowledgments}
This work was supported by the National Natural Science Foundation of China under Grants No. 11890714, No. 11835002, No. 11421505, No. 11522547 and No. 11375251, the Key Research Program of the Chinese Academy of Sciences under Grant No. XDPB09. X. L. Zhao was supported by the Chinese Government Scholarship of Chinese Scholarship Council under CSC Grant No. 201804910796.
\end{acknowledgments}

\end{document}